\documentclass{PoS}
\usepackage{graphicx}% Include figure files

\usepackage{dcolumn}% Align table columns on decimal point
\usepackage{bm}% bold math
\usepackage[numbers]{natbib}
\usepackage{graphicx}
\usepackage{amssymb}
\usepackage{epstopdf}

\def  \be   {\begin{equation}}
\def  \ee   {\end{equation}}
\def  \beq  {\begin{eqnarray}}
\def  \eeq  {\end{eqnarray}}

\title{Experimental biases on the heliospheric contribution to the observed TeV cosmic ray anisotropy}

\ShortTitle{Experimental biases on TeV cosmic ray anisotropy}

\author{\speaker{J. C. D\'iaz--V\'elez} \\
Wisconsin IceCube Particle Astrophysics Center (WIPAC), University of Wisconsin, Madison, WI 53703 \&
Centro Universitario de los Valles, Universidad de Guadalajara,  C.P. 46600, Ameca, Jalisco, M\'exico \\
E-mail: \email{juan.diazvelez@alumnos.udg.mx}
}
\author{Paolo Desiati \\
Wisconsin IceCube Particle Astrophysics Center (WIPAC), University of Wisconsin, Madison, WI 53703 \\
E-mail: \email{desiati@icecube.wisc.edu}
}

\abstract{The arrival direction distribution of cosmic ray particles observed on Earth is shaped by the cumulative effects of their galactic source locations and of trajectory bending in the turbulent interstellar magnetic field. Coherent magnetic structures are expected to disrupt particle trajectories and their observed distribution, as well. The heliosphere, the large magnetic bubble generated by the sweeping effect of solar wind on the local interstellar plasma, strongly affects the TeV cosmic ray particles detected on Earth. By unfolding the heliospheric influence on the observed anisotropy, it is possible to determine the pitch angle distribution of cosmic rays in the interstellar medium. This information makes it possible to study in detail the global diffusion properties of TeV cosmic rays in the Galaxy. However, observational blindness to key features of the cosmic ray arrival direction distribution may lead to biases in the determination of the heliospheric influence and of the interstellar CR distribution. Any inference of galactic TeV CR diffusion properties must carefully account for local propagation phenomena and observational limitations.}

\FullConference{36th International Cosmic Ray Conference -ICRC2019-\\
		July 24th - August 1st, 2019\\
		Madison, WI, U.S.A.}

\begin{document}

\section{Introduction}
It's been over a decade since large ground-based experiments have began providing detailed observations of cosmic rays (CR) arrival direction distributions, in both the northern and the southern hemispheres. It was found that CR flux has an energy dependent anisotropy with amplitude of order 10$^{-4}$-10$^{-3}$ in the energy range of TeV-PeV~\citep{nagashima_1998,hall_1999,amenomori_2005,Amenomori:2006bx,guillian_2007,Milagro:2008nov,abdo_2009,aglietta_2009,munakata_2010,IceCube:2010aug,Abbasi:2011ai,MINOS:2011icrc,IceCube:2012feb,IceCube:2013mar,ARGO-YBJ:2013gya,HAWC:2014dec,bartoli_2015,icecube2016,tibet2017,bartoli2018,hawclsa2017}.
The anisotropy is quantified by determining the relative difference of the number of CR induced events collected in a given direction, with respect to that expected from an isotropic CR flux. Due to the small amplitude of TeV CR anisotropy, the determination of an experiment's response to isotropic flux is performed by averaging the actual event rate variations in time and viewing angle. Due to the limited field of view at any given time, a fixed direction on the sky is only visible for a fraction of a day, i.e. a complete rotation across the celestial sphere, thus affecting the correct CR event all-sky averaging. With appropriate reconstruction methods, it is possible to iteratively account for the sidereal day integrated field of view and properly compensate for the angular dependency of the experiment's exposure \cite{ahlers_method}. However, estimations of an Earth-based experiment's exposure using actual data inevitably requires averaging along declination bands, which washes out any North-South dependency of the observed CR anisotropy.

Ground-based CR experiments typically have wide energy response and poor energy resolution, mainly because CRs are indirectly detected via the secondary particles produced by their interaction with Earth's atmosphere. Since only a fraction of particles in extensive air showers are detected, statistical fluctuations play a crucial role in determining the energy resolution of the primary incident particles. In addition, ground-based experiments have very poor resolution in primary CR mass, aggravated by the fact that unfolding CR particle type is limited by the extrapolations of high energy hadronic interaction models in the forward region~\cite{Antoni_2001}. CRs have a complex energy-dependent mass composition~\cite{PhysRevD.98.030001} and while direct experiments (e.g., balloon or satellite-borne) can identify each primary particle, at ground level it is possible only to estimate mass on a statistical level. Therefore, any sky map of CR arrival direction distribution comprises primary particles of various masses within a relatively wide energy range, typically indicated with its median or average value. In other words, CR anisotropy observations include particles in a wide rigidity range.

The relative intensity sky map of CR flux is characterized by a complex topology, that can be interpreted using a spherical harmonic expansion. This techniques makes it possible to study the CR flux distribution in terms of its dipole, quadrupole and higher multipole components. The corresponding power spectrum describes how anisotropy is topologically distributed. 
Recently, the HAWC gamma-ray observatory (located in Mexico at latitude of 19$^{\circ}$N) and the IceCube neutrino observatory (located at the geographic South Pole at 90$^{\circ}$S) have combined their collected data to produce the first nearly full-sky map of CR anisotropy at a median energy of 10 TeV \cite{Abeysekara_2019}. The advantage of the joint data analysis is to compensate for the strong correlation between spherical harmonic components arising from partial time-integrated field of view of each individual ground-based experiment. The observations show that 99\% of the angular power is mostly concentrated on the large scale anisotropy, i.e. the dipole, quadrupole and octupole components ($\ell$ = 1,2,3). Only about 1\% is observed in medium/small scale structures ($\ell >$3).

CR particles with rigidity of 10 TV in a magnetic field of 3 $\mu$G, have maximum gyroradius of about 730 AU, which is the same order of magnitude of the transverse size of the heliosphere~\cite{Pogorelov:2009may,Pogorelov:2013jul,1742-6596-719-1-012013}, the comet-shaped magnetic bubble surrounding the solar system, formed by the interaction of the highly ionized solar wind and the flow of partially ionized local interstellar medium (LISM). The heliospheric influence on TeV CR particle trajectories is expected to significantly disrupt their arrival direction distribution on Earth, and is characterized by chaotic behavior (paper in preparation). The observed TeV CR anisotropy, therefore, is shaped by the long-term diffusion properties across the interstellar medium and by the short-range non-diffusive effects due to turbulence and the heliosphere. In order to determine the properties of CR galactic diffusion, it is necessary to understand, characterize and unfold the heliospheric influence~\cite{Zhang_2014}. Not only such study requires the state of the art modeling of the heliosphere, but the most unbiased observations, as well.

The full-sky map anisotropy observation, provides a novel unbiased view of the 10 TeV CR flux which makes it possible, for the first time, to study phenomenological scenarios at the origin of the anisotropy. However, the blindness to the North-South components of the anisotropy, i.e. to all $m=0$ terms in the spherical harmonic expansion, may have undesired effects leading to significant mis-interpretations of the results.

In this work, numerical trajectory integration of CR particles propagating across the heliospheric magnetic field are performed, taking into account the wide energy response of typical ground-based experiments and their poor mass resolution, and by performing the same sky map reconstruction as done with real data \cite{ahlers_method}. The numerical calculations are performed by back-propagating anti-particles from Earth's location outward into the LISM, and assuming that trajectories can be time-reversed. The scope of this work is to study how the heliosphere influences CR arrival direction distribution on Earth depending on their distribution in the LISM, and how the North-South blindness biases the determination of such distribution.

\section{Numerical Calculations} 
\begin{figure}[!ht]
\centering
\includegraphics[width=.6\columnwidth]{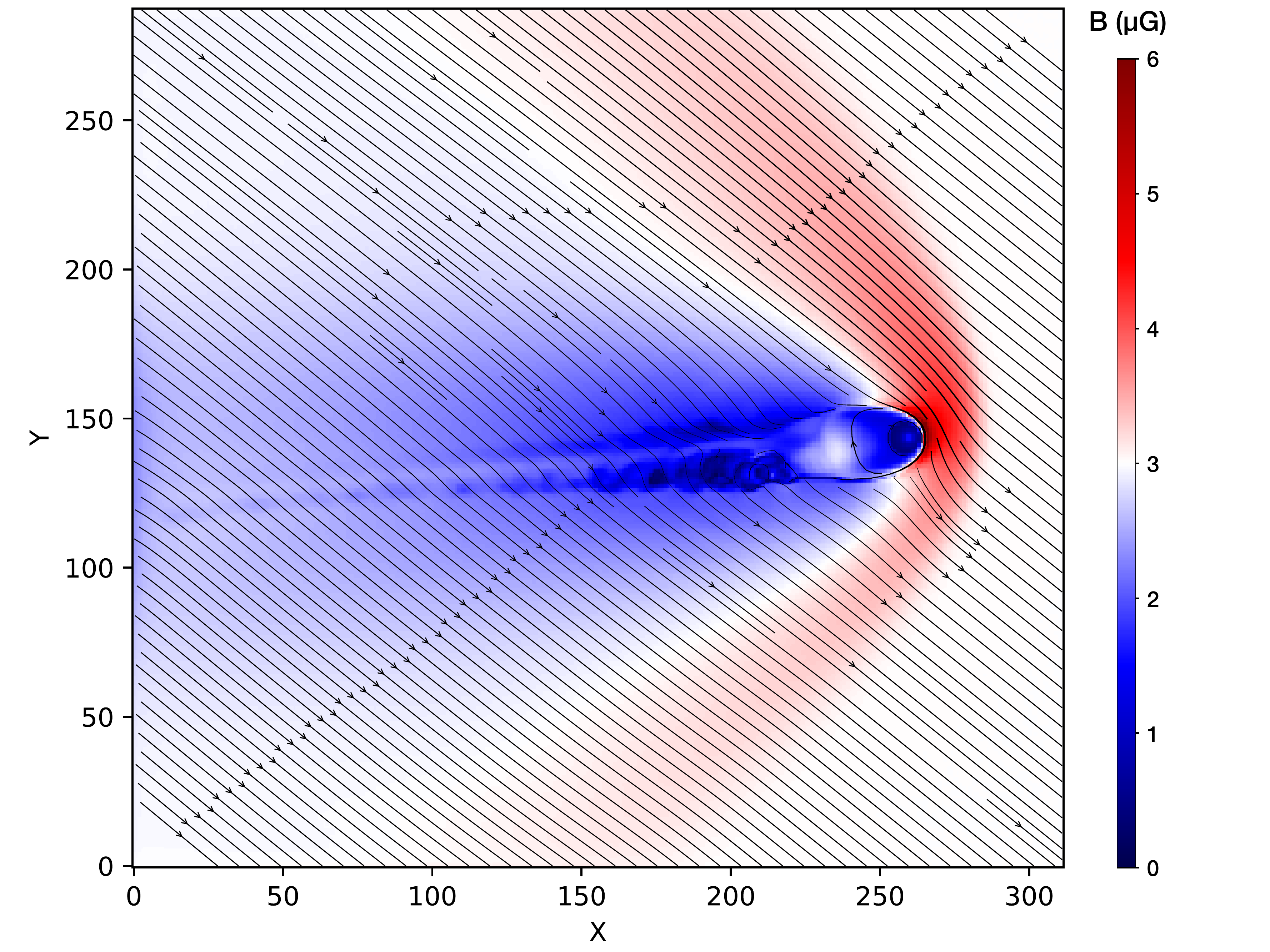}
\caption{Meridional projection of the heliospheric magnetic field model described in~\cite{pogorelov_2013}. The original simulation box dimension is 320$\times$280$\times$280 grid points (20 AU per grid point).}
\label{fig:heliofield}
\end{figure}

Trajectories are calculated by numerically integrating
the following set of 6--dimensional ordinary differential equations
\begin{equation}\label{eq:motion}
\frac{d\mathbf{p}}{dt} = q \left(\mathbf{v}\times\mathbf{B} \right),
\,\,\,\,\,\,\,\,\,\,\,\,
\frac{d\mathbf{r}}{dt} = \mathbf{v},
\end{equation}
describing the Lorentz force exerted by the magnetic field $\mathbf{B}$ on particles with velocity $\mathbf{v}$, where $\mathbf{r}$ is their position vector and $\mathbf{p}$ the momentum. In this work, electric fields are neglected (see ref. \citep{Desiati_2014} for more details on the numerical calculations). 
The heliospheric magnetic field implemented in the numerical calculation is that used in L\'opez-Barquero et al.~\cite{L_pez_Barquero_2017}. It makes use of ideal Magneto-Hydrodynamic (MHD) treatment of ions and of a kinetic multi-fluid description of neutral interstellar atoms penetrating into the heliosphere~\cite{pogorelov_2013}. Figure~\ref{fig:heliofield} shows the meridional projection of the model.
For the present study, the original simulation box is extended into a sphere with radius 50000 AU and centered at the location of the Sun. In the extension of the simulation box, a uniform magnetic field with intensity 3 $\mu$G and with the same direction as in the simulation is assumed. Such a model of the heliosphere is suitable for studying its effects on CR particles with gyroradius comparable to or larger than the heliosphere's size.

The equations of motion (\ref{eq:motion}) are numerically integrated using the explicit 4$^{th}$ order Runge-Kutta integration method. An adaptive time step size was adopted in the calculations, with tolerance level $\epsilon = 10^{-12}$ to keep truncation errors sufficiently small (see Ref. \citep{Desiati_2014} for more discussion on numerical accuracy). The maximum integration time was set to $t_{max} = 330$ years in physical units.
In order to account for the wide energy response of ground-based experiments, nine trajectory data sets were numerically integrated, corresponding to rigidities of $R \in \left[3, 6, 10, 18, 30, 60, 100, 180, 300\right]$ TV.
For each set, a total of 12,288 anti-proton trajectories are integrated back in time starting from the Earth's location (assumed to coincide with the Sun in the model's scale) and initial momentum direction corresponding to each pixel in a HealPix grid with Nside=32. The numerical integration proceeds until trajectories reach the sphere at radius of 50,000 AU. Such distance is always reached well within $t_{max}$ for all the sets. Each integrated data set represents a mapping between a uniformly distributed particle distribution on Earth and that in the LISM as shaped by the LIMF and the heliosphere.

\section{CR Anisotropy through the Heliosphere}\label{sec:helio}
The experimental response to median energy of 10 TeV, from a hypothetical ground-based experiment, is represented with a gaussian weighting function of $\log R$ centered at $\log R = 1$ (10 TV) and with $\sigma_{\log R}$ = 0.5. All trajectory sets are combined into one using this weighting function, and using the mass composition from \citep{gst}, thus representing the experimental smearing in the estimation of CR primary particle energy and the additional dispersion represented by the fact that 10 TV proton has 10 TeV energy while 10 TV Fe nucleus has 260 TeV energy.

The experimental response by a pair of ground-based experiments (one in the northern and one in the southern hemisphere) can be represented using an arbitrary angular response function for each of them, since the iterative reconstruction method used to determine the map in relative intensity properly compensates for that distribution (see Refs. \citep{ahlers_method,Abeysekara_2019}). In this study we assume the angular response function as in Ref. \citep{ahlers_method}: 
\begin{equation}
   \mathcal{A}(\theta,\varphi) \propto \cos\theta\left[1 + A\sin\theta\sin^2(\varphi-\varphi_0)\right]\,,
\end{equation}
in local coordinates $\theta$ and $\varphi$.
The true map in relative intensity of CR flux is then determined for the combined trajectory set. After applying the reconstruction method, the resulting sky map is equivalent to the true map, with the exception of missing $m=0$ terms in each spherical harmonic component (i.e. the North-South blindness).
In order to evaluate the effect such a North-South blindness has in the mapping of CR particle trajectories between Earth and the LISM, the reconstruction method is applied in two distinct procedures. In the first, the CR particle arrival distribution on Earth is determined assuming that their pitch angle distribution in the LISM follows a dipole function aligned to the LIMF with amplitude of $\delta_T = 10^{-3}$; in the second procedure, the distribution in the LISM is determined assuming the numerically calculated and experimentally determined CR distributions on Earth.

\section{Distribution of CRs on Earth}\label{earth_dist}
\begin{figure}[!ht]
\centering
\includegraphics[width=.7\columnwidth]{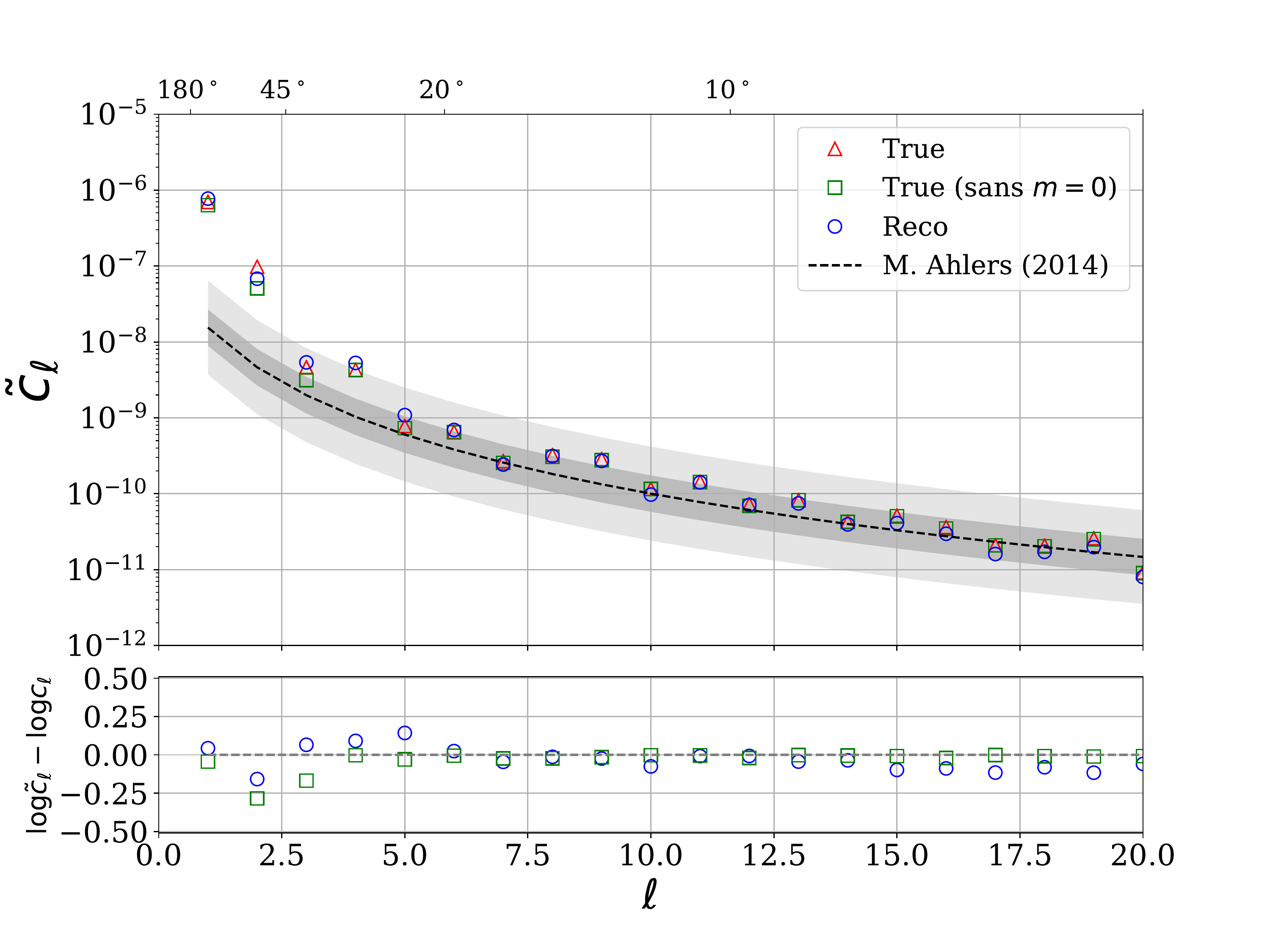}
\caption{The angular power spectrum measured at Earth. While most of the angular power is  concentrated on the large scale anisotropy components ($\ell$ = 1,2,3) a hierarchical ordering of higher $\ell$ modes is present and preserved in the reconstructed map.}
\label{fig:powerspectrum}
\end{figure}
Assuming a simple dipolar pitch angle distribution of CRs in the LISM with amplitude $\delta_{LISM}$ and aligned along the LIMF, it is possible to determine their arrival direction distribution on Earth after propagation through the heliosphere using the combined trajectory set. Decomposing the reconstructed relative intensity map in spherical harmonic components, provides a breakdown into the dipole, quadrupole and higher order multipole contributions arising from the interactions with the heliosphere (Fig. \ref{fig:powerspectrum}). The distribution of angular power for $\ell >4$ agrees with the hierachical model for large $\ell$ described by the relation $C_\ell \propto 1/(2\ell+1)(\ell +2)(\ell+1)$ from Ref.~\citep{Ahlers_2014}.

The TeV CR anisotropy at the largest scale (i.e. $\ell = 1, 2, 3$, that holds most of the power) is significantly affected by the heliosphere's influence, in that it re-distributes the power in a way that depends on the heliospheric magnetic field.
The missing $m=0$ terms due to the North-South blindness has the strongest effect on the reconstructed dipole, where only the horizontal component $\delta_H$ is experimentally available. However, the impact on higher order multipole terms diminishes with higher $\ell$, since only 1 of the $2\ell+1$ components is missing and the 2D sky map features are retained with higher accuracy the higher the $\ell$.

\begin{figure}[h!]
\centering
\includegraphics[width=.7\columnwidth]{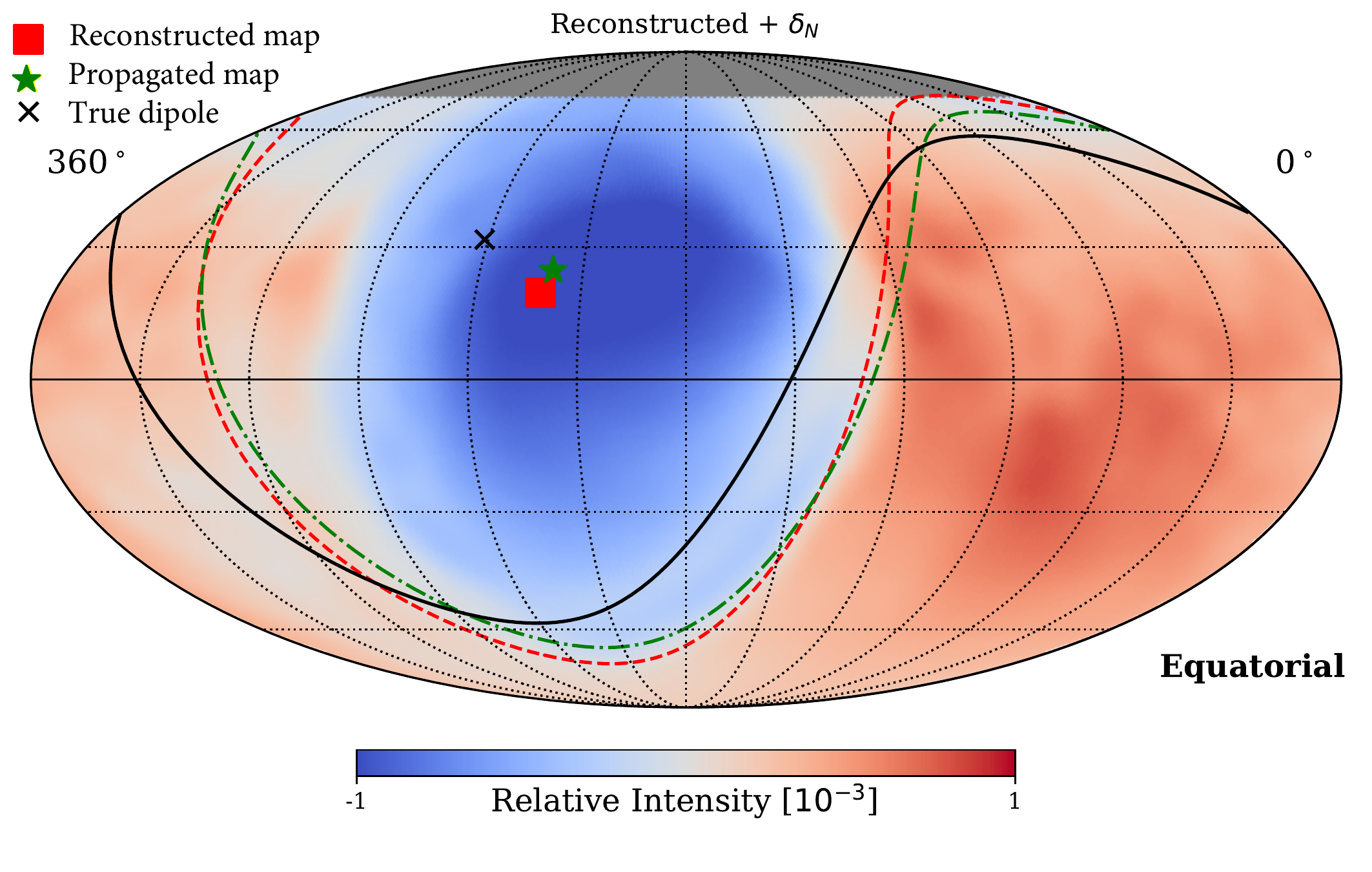}
\caption{The reconstructed map of the 10 TeV combined sample after propagation. The direction of the LIMF is indicated by the X and the corresponding magnetic equator (i.e. the plane perpendicular to the uniform LIMF passing by the Earth) is shown with a solid curve. The inferred direction obtained by fitting a circle to the boundary of large-scale excess and deficit regions (dot-dashed green curve) from the true propagated map is indicated with a green star. The equivalent fit (shown as a dashed red curve) for the reconstructed map is indicated with red square.}
\label{fig:recomap}
\end{figure}

Although the reconstructed dipole contribution lost its vertical component, and therefore its true direction on the celestial sphere, the higher multipole terms have retained more spatial information, providing an overall preservation of the true features in the CR distributions. To show this, a circle fit on the regions in the sky with the largest gradient in relative intensity is performed on the reconstructed sky map. This is the same method used in Ref.~\cite{Abeysekara_2019}. The center of the best circle fit is found to be within a few degrees from the fit result performed with the true sky map distribution and about 10$^{\circ}$ from the direction of the LIMF, as shown Figure~\ref{fig:recomap}.
Therefore, unlike the reconstructed dipole component $\delta_H$, the overall reconstructed sky map still retains the ordering with the LIMF. Assuming that the dipolar CR pitch angle distribution in the LISM is aligned with the LIMF, this makes it possible to estimate the missing North-South dipole component $\delta_N$ of the CR anisotropy. Figure~\ref{fig:recomap} shows the reconstructed sky map of CR relative intensity with the North-South dipole component included.
Comparing the dipole amplitude of the propagated map $\delta_{P}$ 
with the reconstructed dipole amplitude 
$\delta_R = \sqrt{\delta_H^2 + \delta_N^2}$ we find that it agrees to within $1\%$ and in particular that the estimate of $\delta_N$ agrees to within 2\% of the true value.

\section{Distribution of CRs in the LISM}
\begin{figure}[h]
\centering
\includegraphics[width=.7\columnwidth]{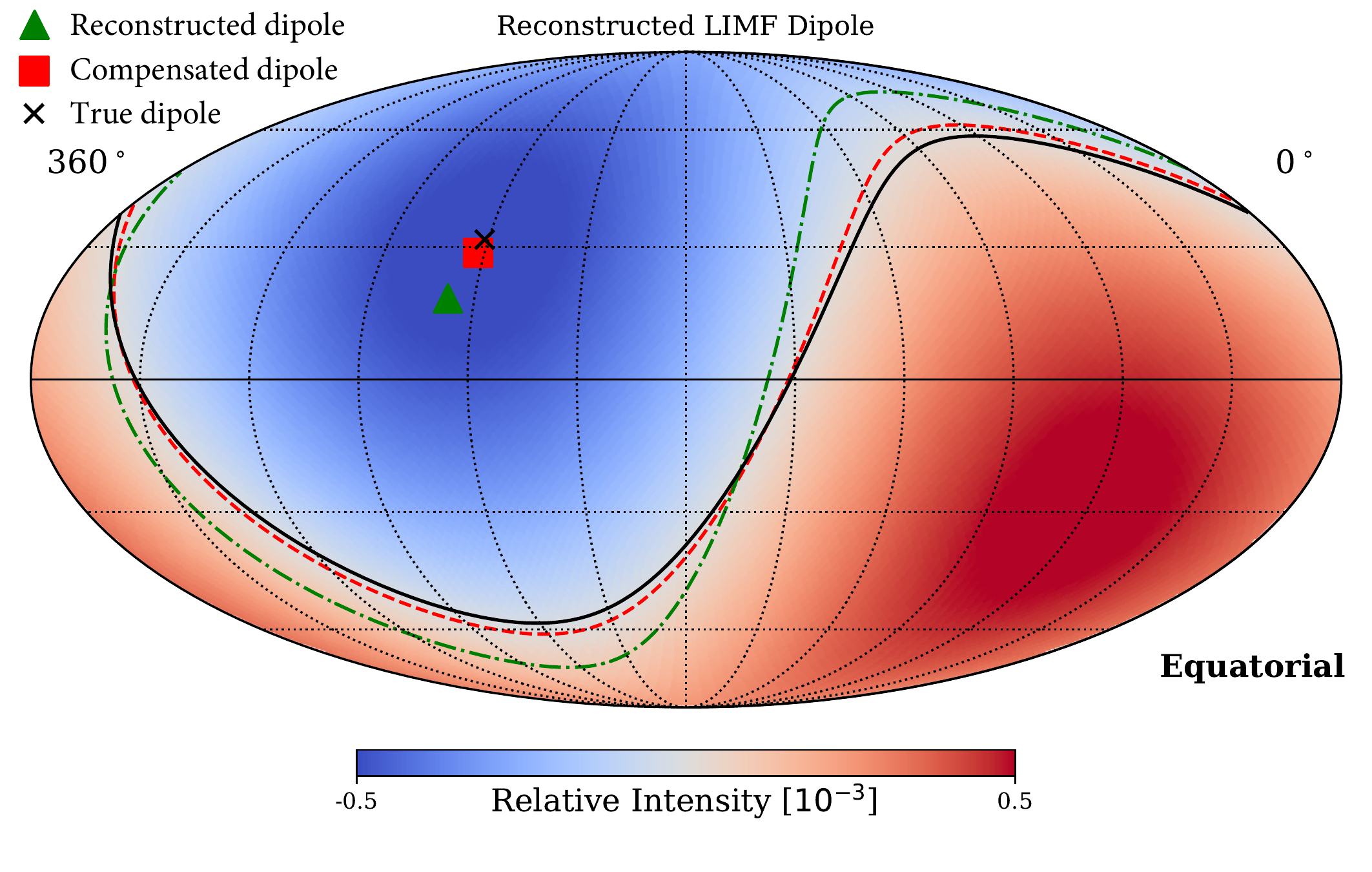}
\caption{The reconstructed map of the 10 TeV combined sample after propagation is remapped from recorded trajectories outside the heliosphere. The true direction of the LIMF dipole is indicated by the X. The inferred dipole direction obtained from backtracking particles is indicated with a  red square, in the case of the $\delta_N$-compensated map, and a green triangle, for the reconstructed map without $\delta_N$-compensation.
}
\label{fig:dipolediff}
\end{figure}

 In order to determine the TeV CR distribution in the LISM, it is necessary to account for the heliosphere and subtract its influence, as done in Ming et al. (2014). 
In order to do this we reverse the procedure described in Sec.~\ref{earth_dist} by re-weighting pixels in the reference system of the LISM using the distribution obtained from the reconstructed map in the reference system of Earth (with and without $\delta_N$ compensation). We compute a superposition of nine maps (corresponding to each of the nine different rigidities $R=$3--300 TV) weighted according to the distribution described in Sec \ref{sec:helio}. Each map is generated from the pixel mapping obtained from back-propagation for a given rigidity $R$. Fig. \ref{fig:dipolediff} shows the reconstructed dipole obtained by remapping the dipole components of the distribution Fig. \ref{fig:recomap}. The inferred dipole direction obtained with and without compensation for the estimated $\delta_N$ dipole. The fits are located at an angular distance of $3^\circ$ and $14^\circ$ away from the true LIMF dipole respectively. This difference is comparable with uncertainties due to energy and mass resolution.

\section{Conclusions and outlook}
As discussed in Ref. \citep{Desiati:2012jun} and numerically evaluated in Refs.~\citep{Schwadron2014} and \citep{Zhang_2014}, the heliosphere influences the arrival distribution of TeV-scale CRs. The strongest effect is expected to be caused by magnetic structured with spatial scale comparable to the particle's gyroradius. CR particles with rigidity scale of 10 TV, have gyroradius comparable to the spatial scale at which the LIMF drapes around the heliosphere~\cite{desiati_lazarian}. Such CRs are more likely to experience magnetic reflection and produce significant deformation of the initial angular distribution.

We have found that, although the heliosphere lensing effect at 10 TV rigidity is strong, the ordering with the LIMF is preserved and it can be determined with good accuracy by taking into account all the features of the CR anisotropy distribution and assuming the pitch angle distribution in the LISM is aligned to the magnetic field lines as was done in Ref.~\cite{Abeysekara_2019}.

Reconstruction biases such as North-South blindness as well as energy and mass resolution can limit our ability to correctly reconstruct the pitch angle distribution of cosmic rays in the interstellar medium. However, the features described by higher $\ell$-modes contain sufficient information in order to unfold the influence of the heliosphere on the observed cosmic ray anisotropy. The approach of back tracing CRs appears to be much more sensitive to the particular LIMF-heliospheric model. 
 
The results from this study will guide future work on studying the effects of modulating parameters within the model itself, including the effects of heliotail length, of solar cycles, turbulence, and the relative direction of LIMF to that of the heliosphere.

\section*{Acknowledgements}
The authors wish to thank colleagues at WIPAC and
the Department of Astronomy for discussions on cosmic
ray anisotropy, and Nikolai Pogorelov for useful discussions on MHD simulations of the heliosphere.  The authors acknowledge the support from the U.S. National Science
Foundation-Office of Polar Programs and Consejo Nacional de Ciencia y Tecnolog\'{\i}a (CONACyT).

\bibliography{HelioAnisotropy}
\bibliographystyle{JHEP}

\end{document}